\documentclass[journal]{IEEEtran}

\ifCLASSINFOpdf
\else
   \usepackage[dvips]{graphicx}
\fi
\usepackage{url}

\usepackage{graphicx}
\usepackage{multirow}
\usepackage{booktabs} 
\usepackage{amsmath}
\usepackage{amssymb}
\usepackage{orcidlink}
\usepackage{subcaption}  
\usepackage{caption}    

\begin{document}

\title{Audio-based Kinship Verification Using \\Age Domain Conversion}

% \author{Qiyang Sun\orcidlink{0009-0001-9228-4543}, Alican Akman\orcidlink{0000-0002-8010-6897}, Xin Jing\orcidlink{0000-0002-8803-9414}, Manuel Milling\orcidlink{0000-0002-8842-2958}, and Björn W. Schuller\orcidlink{0000-0002-6478-8699} \IEEEmembership{Fellow, IEEE}
\author{Qiyang Sun, Alican Akman, Xin Jing, Manuel Milling, and Björn W. Schuller \IEEEmembership{Fellow, IEEE}
% \thanks{This work was supported in part by .}
\thanks{Qiyang Sun and Alican Akman are with GLAM, Department of Computing, Imperial College London, UK (e-mail: q.sun23@imperial.ac.uk; a.akman21@imperial.ac.uk).}
\thanks{Xin Jing and Manuel Milling are with CHI – Chair of Health Informatics, MRI, Technical University of Munich, Germany (e-mail: xin.jing@tum.de; manuel.milling@tum.de).}
\thanks{Björn W. Schuller is with GLAM, Department of Computing, Imperial College London, UK; CHI – Chair of Health Informatics, MRI, Technical University of Munich, Germany; MDSI – Munich Data Science Institute, Munich, Germany; and MCML – Munich Center for Machine Learning, Munich, Germany (e-mail: bjoern.schuller@imperial.ac.uk).}
\thanks{Code is available at: {\href{https://github.com/sqy991018/ageconversion}{https://github.com/sqy991018/ageconversion}}.}

}

% \markboth{IEEE SIGNAL PROCESSING LETTERS, VOL. , 202}
% {Shell \MakeLowercase{\textit{et al.}}: Bare Demo of IEEEtran.cls for IEEE Journals}
\maketitle

\begin{abstract}
Audio-based kinship verification (AKV) is important in many domains, such as home security monitoring, forensic identification, and social network analysis. A key challenge in the task arises from differences in age across samples from different individuals, which can be interpreted as a domain bias in a cross-domain verification task. To address this issue, we design the notion of an ``age-standardised domain" wherein we utilise the optimised CycleGAN-VC3 network to perform age-audio conversion to generate the in-domain audio. The generated audio dataset is employed to extract a range of features, which are then fed into a metric learning architecture to verify kinship. Experiments are conducted on the KAN\_AV audio dataset, which contains age and kinship labels. The results demonstrate that the method markedly enhances the accuracy of kinship verification, while also offering novel insights for future kinship verification research.
\end{abstract}


\IEEEpeerreviewmaketitle

\section{Introduction}
\label{sec:intro}
\IEEEPARstart{K}{inship} verification (KV) is an active area of research \cite{wang2023survey}, which is defined as the process of using deep learning algorithms to identify whether two individuals are related by blood through the extraction and comparison of information from biometric data \cite{nader2021kinship}. It has strong social significance and applies to a variety of fields, including family security monitoring, rapid forensic identification, and social network analysis \cite{wu2022facial}. The majority of current research has focused on the use of facial images and videos to identify kinship as, in general, genetically related family members exhibit pronounced similarities in terms of appearance \cite{ghahramani2014family}. Corresponding similarities, however, can also be observed in voice characteristics as backed up by a recent study~\cite{wu2019audio}. Nevertheless, in line with the common favouritism of the image modality over the audio modality~\cite{milling2024bringing}, audio-based kinship verification (AKV) has barely been investigated, despite some key advantages of the audio data in the task setting: for instance, in many scenarios, such as phone calls, video footage is unavailable as it is impractical to collect.

Generally speaking, KV requires recognising certain similarities between individuals resulting from a family connection while condoning similarities that result from other domain overlaps, such as age or gender. In many cases, related individuals belong to different domains, e.g., a mother-son pair having different age and gender. This adds to additional challenges with respect to variations in appearance or voice characteristics ~\cite{oscarson2024face}, which can be even increased due to emotional states or differences in environments or recording devices~\cite{hettiachchi2020augmenting,luo2018deep,mzoughi2024review}. 

In an attempt to mitigate these domain biases resulting from different data distributions~\cite{tu2012cross}, one approach is to learn specific latent feature spaces \cite{wang2018cross}. Most prominently, in AKV, the speech characteristics pitch, timbre, tone, and speech rate, to name but a few, vary largely across age groups~\cite{eichhorn2018effects, gisladottir2023sequence}. 
To address countering these domain bias effects, we explore voice conversion techniques in this work to leverage voice recordings of individuals of different ages into the same age domain via generative models. This age-standardised domain refers to a target age range where speech characteristics are unified, minimising the influence of age-related variations in voice. Specifically, young speech data is converted by ageing and projected into a middle-aged standard domain, old speech data is processed by rejuvenation and similarly projected into the middle-aged domain. 
Subsequently, high-dimensional features are extracted and sent to a triplet network \cite{hoffer2015deep} for kinship verification.
Our experiments demonstrate that an age-conversion can be helpful in increasing kinship verification accuracy compared to the baseline consisting of the original recordings from different age domains.

% figure
\begin{figure*}[t]
  \centering
  \includegraphics[width=\textwidth]{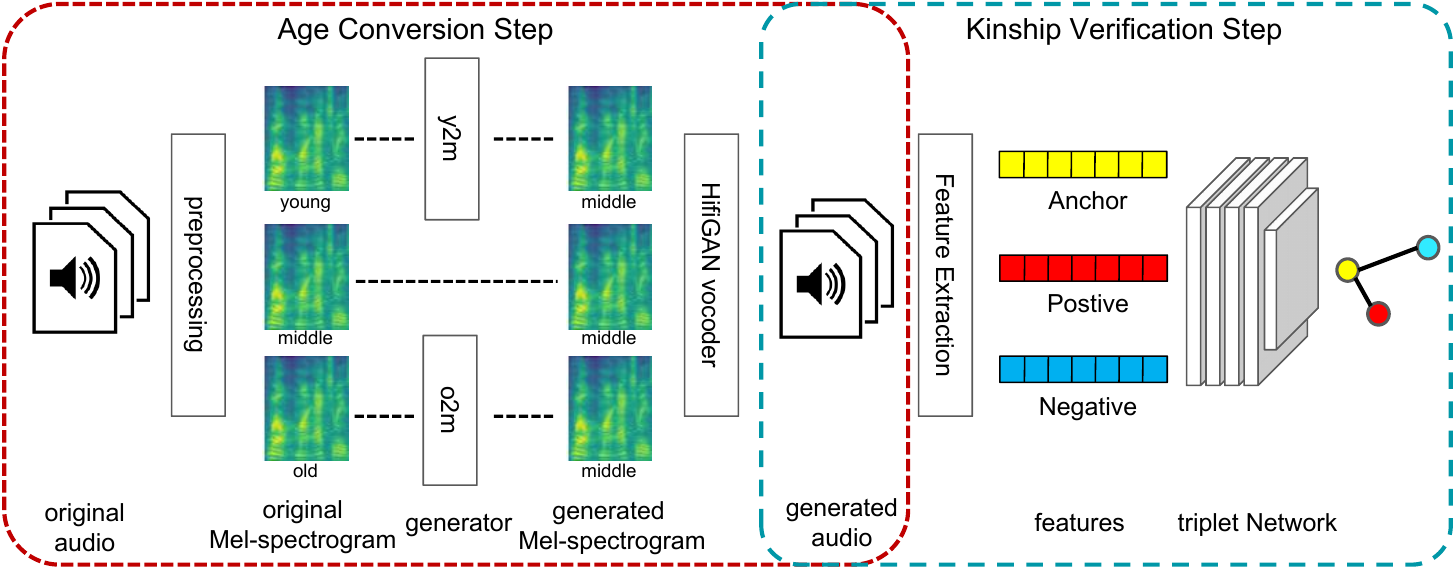}
  \caption{Framework of kinship verification using age voice conversion. }
  \label{fig:framework}
\end{figure*}

\section{Related Work}

\label{sec:related}
\subsection{Voice Conversion}
Voice conversion (VC) is a technique that is commonly used in the field of speech processing and generation to modify and manipulate specific properties of voices \cite{MOHAMMADI201765}. It encompasses the alteration of a speech's gender, age, emotion and timbre \cite{yang2022overview,sisman2020overview}. Through the analysis and synthesis of audio features, VC techniques can be used to generate audio with characteristics that closely align with those of the target \cite{mu2021review, lee2024hear}. It has potential applications across a variety of domains, including personalised voice assistants, voice camouflage, and entertainment \cite{akhter2022analysis}. Earlier VC methods are based on Gaussian Mixture Models (GMMs) and Hidden Markov Models (HMMs) \cite{kobayashi2014regression}, which have been largely replaced by more advanced approaches in recent years. With the development of deep learning techniques, models such as Generative Adversarial Networks (GANs) and Variational Autoencoders (VAEs) are increasingly popular in the voice conversion task \cite{bargum2023reimagining, 904d4bb7b8e546e69c7ad27a7ba6b1db, triantafyllopoulos2023overview}. Models such as ACVAE-VC \cite{kameoka2019acvae}, CycleGAN-VC \cite{kaneko2018cyclegan}, FragmentVC \cite{hledikova2022data},  and StarGAN \cite{he2021improved} demonstrate efficacy in gender conversion and speaker identity conversion experiments. However, age-oriented voice conversion remains a relatively underexplored area. As far as we know, only in \cite{padmini2022age}, the researchers propose a blood-relation-based voice conversion technique, utilising the PSOLA algorithm to adjust prosodic features enabling conversion across different age groups to help speech-impaired individuals generate more natural-sounding voices.

\subsection{Audio Kinship Verification}

Within AKV, most approaches comprise two principal stages: feature extraction and comparison learning \cite{wang2023survey}. The most common audio feature representations are i-vector and x-vector \cite{kefalas2023kan}. i-vector generates low-dimensional feature vectors by mapping based on the GMM model; x-vector employs a deep neural network (DNN) to aggregate frame-level features and generates fixed-length vectors.
Both feature types have their origin in speaker verification~\cite{jakubec2024deep,mohammadi2022weighted}. Subsequently, those features are separately fed into metric learning methods or methods such as Probabilistic Linear Discriminant Analysis (PLDA) for the determination of kinship \cite{wu2022facial, jakubec2021deep}.
 
To address the issue of cross-domain effects in the image-based KV, some studies have sought to mitigate the impact of domain bias on KV under different lighting conditions by projecting data into colour space to extract manual features \cite{nader2024enhanced}. In a separate study, researchers aim to mitigate the age-based domain difference issue in KV. They employ GANs to convert image samples of the elderly into those of the young and used SDM-Loss to guide the deep features learning \cite{wang2018cross}, which also motivates our study for AKV. 
To the best of our knowledge, this work presents the first approach applying generative models to overcome age bias within the field of AKV.

\section{Methodology}
\label{sec:method}
This section details the design of our AKV experiments. An overview of our overall methodology is given in Figure \ref{fig:framework}.

\subsection{Data}
\label{subsec:data}
We use the KAN\_AV dataset in this study, which comprises a substantial corpus of audio and video data collected in a natural setting, with a particular focus on the analysis of faces and voices. The data is collected from publicly accessible video sources, including cinematic clips, oratory presentations, and interviews. The dataset consists of over 28,000 audio and video clips, with a total duration of approximately 98.5 hours, encompasses 970 individuals across a diverse age range, from 3 to 100 years old. 645 individuals have been assigned kinship labels, which fall into one of seven categories of first-degree kinship: The kinship labels fall into one of seven categories of first-degree kinship: father and son (FS, 99 pairs), father and daughter (FD, 82 pairs), mother and son (MS, 46 pairs), mother and daughter (MD, 64 pairs), brother and sister (BS, 93 pairs), brother and brother (BB, 111 pairs), or sister and sister (SS, 56 pairs) \cite{kefalas2023kan}. The dataset is initially divided into three age-based groups: a young group ($<$35 years old, 8032 instances), a middle-aged group (35-55 years old, 12170 instances), and an old group ($>$55 years old, 7798 instances).

We use the audio dataset with 16\,kHz sample rate. For preprocessing, the DC offset of the audio is eliminated, and the audio is converted to single channel. We extract a Mel-spectrogram with a hop length of 256, a window length of 1024, 80 Mel bands in in the frequency range from 0\,Hz to 8,000\,Hz and an FFT size of 1024. We employ the Slaney norm for normalisation and Mel frequency scale conversion. During the training phase, the audio length is adjusted to 64 frames (approximately 1\,s). The generated Mel-spectrogram is then standardised according to the mean and standard deviation of the middle-aged group data. The processed data is used to train our voice conversion model.

\subsection{Age Conversion}

Our age VC model is based on the CycleGAN-VC3 architecture \cite{kaneko2020cyclegan}. CycleGAN aims to achieve unsupervised image-to-image conversion, or in this case, spectrogram-to-spectrogram conversion between two different domains through bi-directional cyclic consistency loss, which has shown excellent performance in voice conversion tasks. The CycleGAN-VC3 model introduces a Time-Frequency Adaptive Normalisation (TFAN) module, which is an extension of the traditional instance normalisation, allowing for a more fine-grained tuning of the features in the time-frequency dimension while preserving the information of the source spectrogram.

The generator first converts the input Mel-spectrogram into 128 channels via a 2D convolutional layer, followed by two downsampling layers, which progressively halve the spatial resolution while increasing the number of channels to 256. During this process, the TFAN module is applied to maintain consistency across time and frequency. Next, six residual blocks are used to extract deep voice features, and a 1D convolution is applied to convert the features back to a 2D representation. During the upsampling phase, the model restores the spatial resolution and increases the number of channels to 1024, with another application of the TFAN module to ensure feature retention. Finally, a 2D convolution generates the output Mel-spectrogram. The discriminator retains the PatchGAN architecture proposed in \cite{kaneko2020cyclegan}.

The generator and discriminator loss functions consist of LSGAN loss, cycle consistency loss, and identity loss. The LSGAN loss improves training stability by replacing the traditional cross-entropy loss with a least-squares objective \cite{mao2017least}. The cycle consistency loss ensures that after mapping samples from one domain to another and back, the original data is preserved, enforcing consistency across domain transformations \cite{zhu2017unpaired}. The identity loss \cite{kaneko2018cyclegan} is used to preserve input characteristics when the input already belongs to the target domain, ensuring that the generator does not alter the data unnecessarily. 
The total loss function is defined as:

\begin{equation*}
\begin{split}
\mathcal{L}(G, F, D_X, D_Y) = \mathcal{L}_{\text{GAN}}(G, D_X) + 
\mathcal{L}_{\text{GAN}}(F, D_Y) + \\
\lambda_{\text{cycle}} \mathcal{L}_{\text{cycle}}(G, F) + \lambda_{\text{identity}} \mathcal{L}_{\text{identity}}(G),
\end{split}
\end{equation*}
where \(\mathcal{L}_{\text{GAN}}(G, D_X)\) and \(\mathcal{L}_{\text{GAN}}(F, D_Y)\) represent the LSGAN losses for generators \(G\) and \(F\) against discriminators \(D_X\) and \(D_Y\). \(\mathcal{L}_{\text{cycle}}(G, F)\) represents the cycle consistency loss. \(\mathcal{L}_{\text{identity}}(G)\) represents the identity loss. \(\lambda_{\text{cycle}}\) and \(\lambda_{\text{identity}}\) are weighting factors for the cycle consistency loss and identity loss, set to 10 and 5, respectively, with \(\lambda_{\text{identity}}\) progressively decaying in the later stages of training.

We use the Adam optimiser with $lr=0.0002$ and the discriminator with $lr=0.0001$ in our experiments.

\begin{table*}[t]
\caption{Kinship verification results (accuracy \%), where ``Baseline TripletNet" is the model in \cite{kefalas2023kan}; ``Optimised TripletNet" is our method; ``Original Dataset" refers to the initial KAN\_AV dataset, while ``Generated Dataset" refers the data processed through voice conversion. Abbreviation of kinship corresponding to Section \ref{subsec:data}. The numbers in brackets represent the weight (\%) of the kinship. ``Overall" refers the weighted accuracy. We highlight the best result in each category. }
\label{table:acc}
\centering
\begin{tabular*}{\textwidth}{@{\extracolsep{\fill}} c c c c c c c c c c c} 
 \toprule
Accuracy [\%] & Method & Features & Overall & BB (11) & SS (5) & BS (14) & FD (28) & FS (19) & MD (22) & MS (1) \\ 
 \midrule
 \multirow{6}{*}{Original Dataset}
 & \multirow{3}{*}{Baseline TripletNet} 
 & i-vectors & 64.2&61.6&72.4&64.9&62.1&65.0&65.2&69.9\\
 &  & x-vectors & 63.6 & 61.1 & 68.4 &65.0 & 60.4 & 65.5&  65.2& 65.8\\
 &  & Wav2Vec & 66.7&68.8 & 61.1 &68.3 &66.3& 65.9& 67.5 &66.9 \\
\cmidrule{2-11}
 & \multirow{3}{*}{Optimised TripletNet} 
  & i-vectors & 66.1 & 63.0& 71.8&66.8&65.1&66.4&66.8&70.5 \\
 &  & x-vectors & 63.4 & 63.0&64.8 & 60.3 &59.4 & 67.6& 65.2& 64.3\\
 & & Wav2Vec &69.8 & \textbf{69.2} &\textbf{76.4}&68.6&70.4& \textbf{71.1}&67.6&70.8\\
 \midrule
 \multirow{6}{*}{Generated Dataset} 
 & \multirow{3}{*}{Baseline TripletNet} 
 & i-vectors & 65.1 & 65.0 &58.8& 59.7& 65.6& 64.4& 67.2& 64.6\\
 &  & x-vectors &65.7  &66.3&67.8 &64.0&63.0 &66.8 & 67.9 &70.1\\
 &  & Wav2Vec &69.3 & 66.4& 66.4&\textbf{74.6}  & 69.8&65.8 &70.3&66.7 \\
\cmidrule{2-11}
 & \multirow{3}{*}{Optimised TripletNet} 
 & i-vectors & 67.6 & 67.8& 67.6&63.0&65.8&70.4& 70.2& \textbf{70.9}\\
 &  & x-vectors &67.9 &67.0&65.9  &68.4&68.1& 68.8& 67.5& 69.6\\
 &  & Wav2Vec &\textbf{71.3} & 67.0& 75.6&72.6& \textbf{74.0} &68.3& \textbf{71.1} & 64.4\\
 \bottomrule
\end{tabular*}
\end{table*}

\subsection{Feature Extraction}

For the AKV task, we extract the i-vector and x-vector with the Kaldi toolkit \cite{povey2011kaldi}. In both cases, the speech signal is first converted to MFCC features with a 25\,ms frame length, 10\,ms frame shift, 30 Cepstral coefficients, and 40 Mel frequency filters. 
Voice Activity Detection (VAD) is initially applied to remove non-speech segments. 
For i-vectors, 400-dimensional features are generated based on a Universal Background Model (UBM) with 512 Gaussian components. 
For x-vectors, a pre-trained Time-Delay Neural Network (TDNN) model is used to produce 512-dimensional features.

Finally, we extract 1024-dimensional features (called Wav2Vec below) from the raw audio via the final transformer layer of a pre-trained Wav2Vec 2.0, specifically designed for age and gender recognition tasks \cite{burkhardt2023speech}. 
This model is fine-tuned on several large-scale speech datasets using the Wav2Vec2-Large-Robust architecture, which is capable of capturing high-level semantic information within the speech signal and providing rich acoustic feature representations relevant to age.

% 亲属验证

\subsection{Kinship Verification Model}

We finally employ a triplet network for the AKV task, which projects the extracted features into a new embedding space and minimises the embedding distances between pairs of kinship samples and maximise the corresponding distances between pairs of unrelated samples \cite{yu2020deep}.The model is presented with anchor-positive-negative triplets. When creating the triplets, the anchor and positive samples come from the same familial relationship pair while the negative sample is randomly selected from individuals of the same gender as the positive samples but with no relationship. For example, We apply a custom data partitioning strategy, ensuring the mutual exclusivity of the training, validation, and test sets by dividing the data by individual in a 7:1:2 ratio. After combining, we generate approximately 285,000 sample pairs. the model is trained to optimise embedding distances to distinguish anchor-positive pairs and anchor-negative pairs. During inference, an optimal threshold is determined, which is subsequently used for binary classification on the test set.

The triplet model consists of a two-layer fully connected network. The first layer maps the input features to a 256-dimensional hidden space, applying Batch Normalisation, a ReLU activation function and a dropout layer. The second layer compresses the features further into a 128-dimensional vector space for triplet loss computation. We use the classic triplet loss function, defined in~\cite{hoffer2015deep}. We use SGD with $lr=0.0001 $ and $momentum=0.9$ during training, .

\begin{figure}[t!]  
    \centering
    \begin{subfigure}[b]{0.45\columnwidth} 
        \centering
        \includegraphics[width=\textwidth]{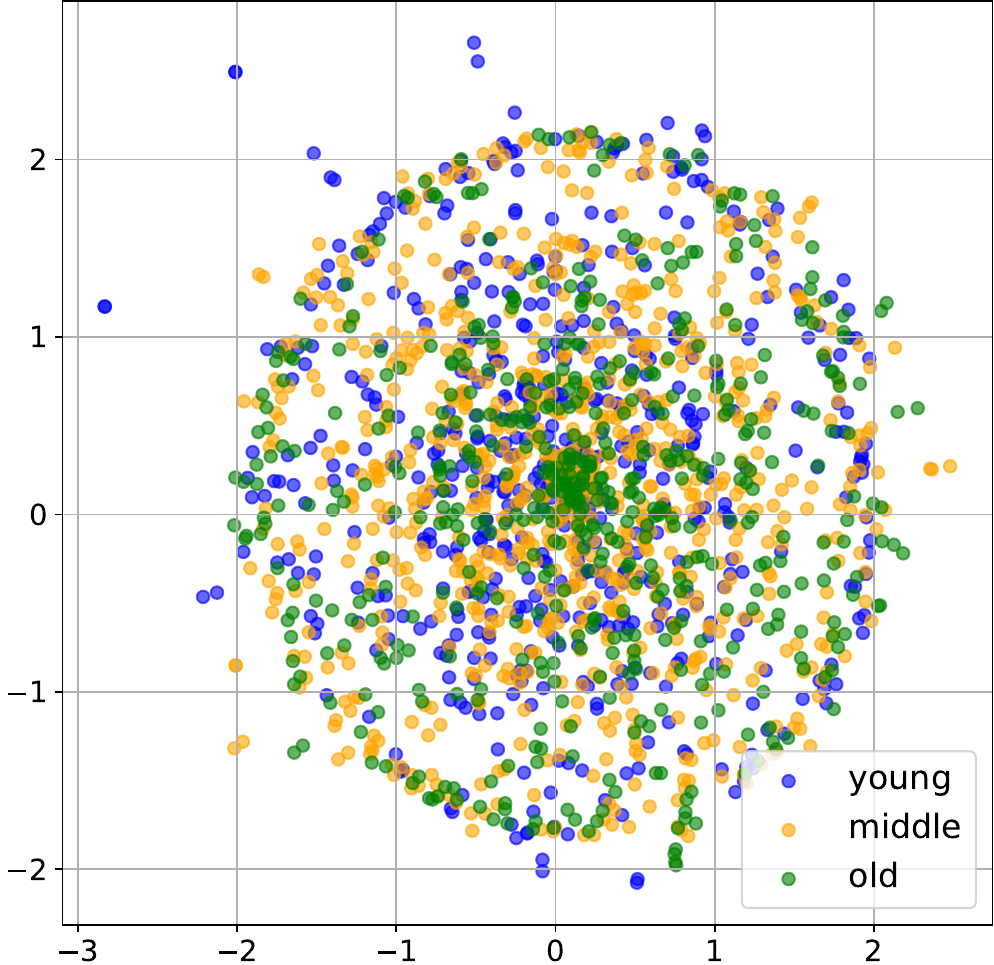}  
        \caption{i-vector original dataset}
        \label{fig:1a}
    \end{subfigure}
    \hfill
    \begin{subfigure}[b]{0.45\columnwidth}
        \centering
        \includegraphics[width=\textwidth]{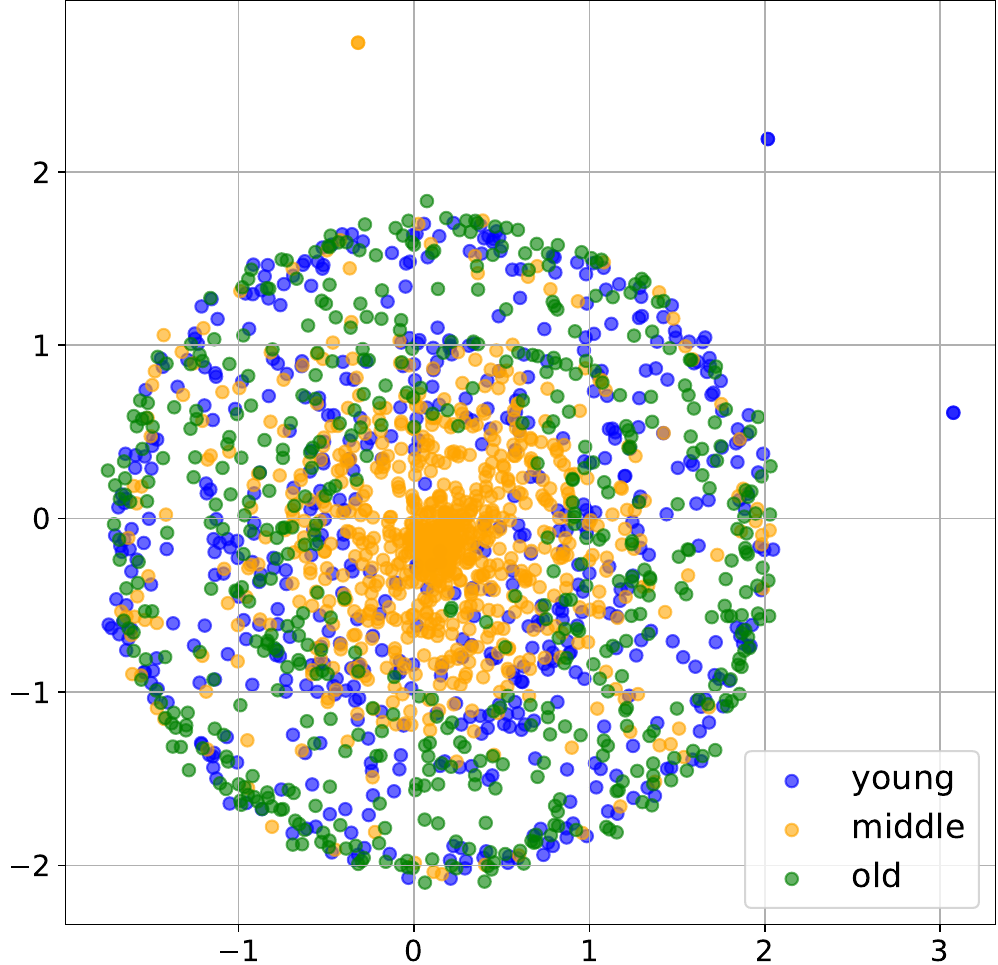} 
        \caption{i-vector generated dataset}
        \label{fig:1b}
    \end{subfigure}

    \begin{subfigure}[b]{0.45\columnwidth}
        \centering
        \includegraphics[width=\textwidth]{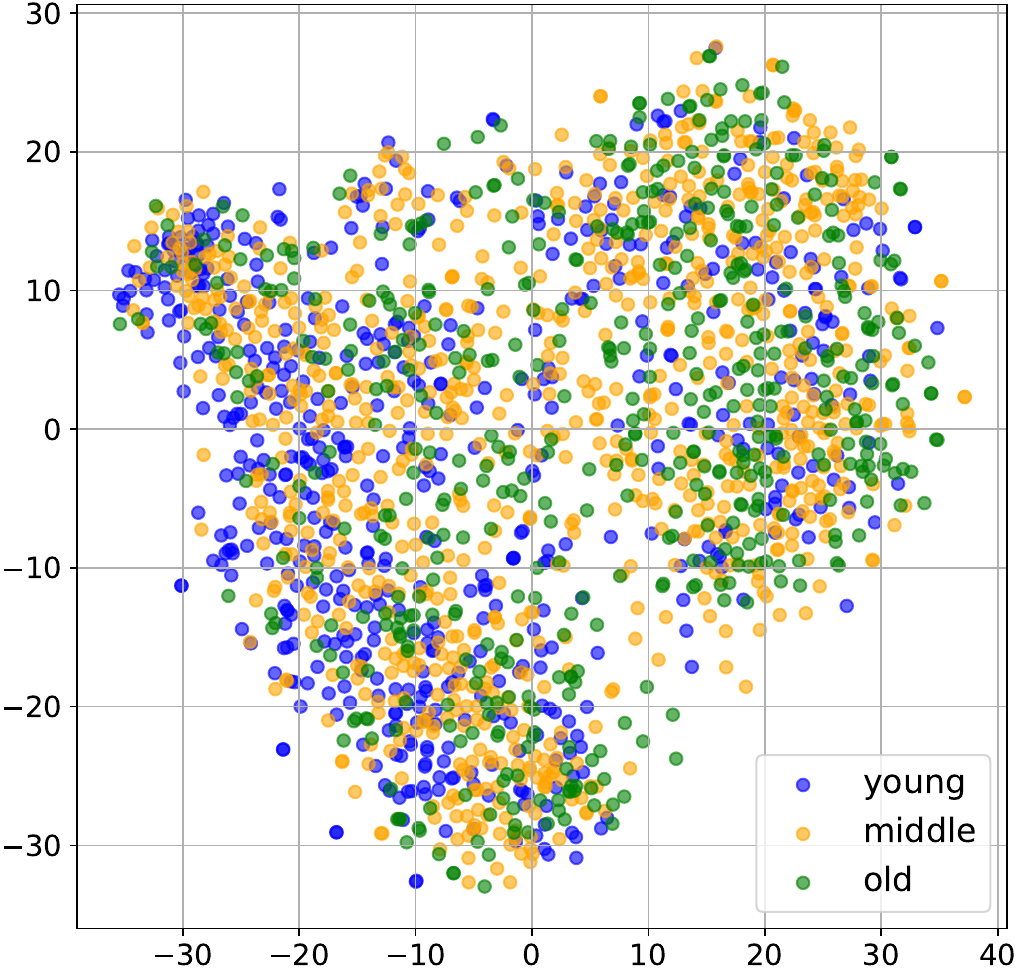}  
        \caption{x-vector  original dataset}
        \label{fig:2a}
    \end{subfigure}
    \hfill
    \begin{subfigure}[b]{0.45\columnwidth}
        \centering
        \includegraphics[width=\textwidth]{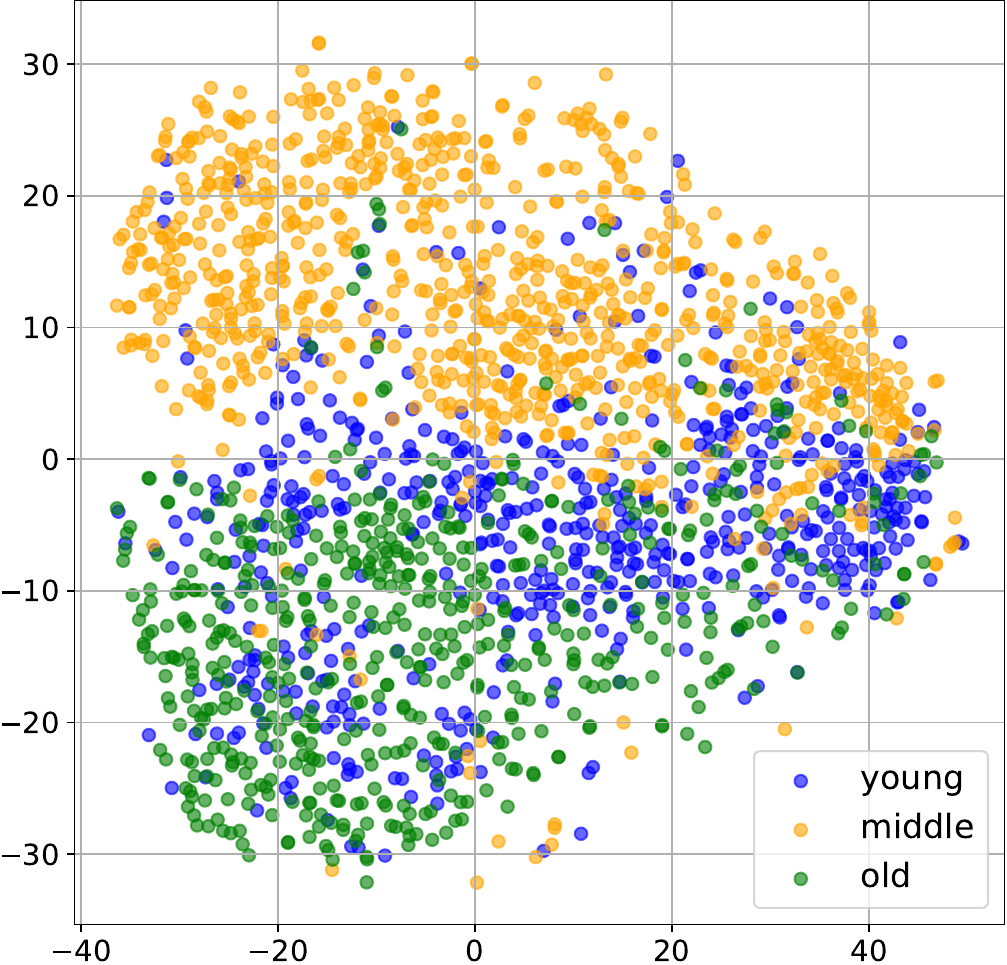}  
        \caption{x-vector  generated dataset}
        \label{fig:2b}
    \end{subfigure}

    \begin{subfigure}[b]{0.45\columnwidth}
        \centering
        \includegraphics[width=\textwidth]{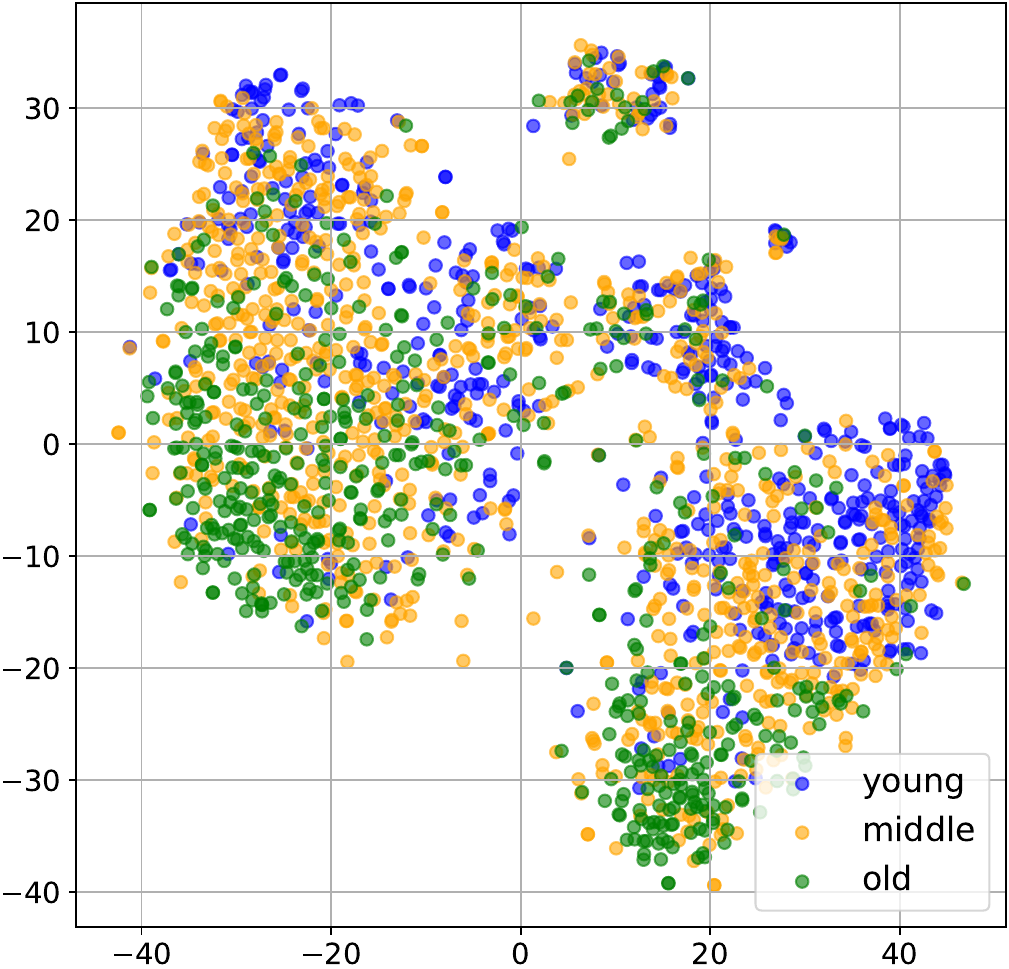}  
        \caption{Wav2Vec  original dataset}
        \label{fig:3a}
    \end{subfigure}
    \hfill
    \begin{subfigure}[b]{0.45\columnwidth}
        \centering
        \includegraphics[width=\textwidth]{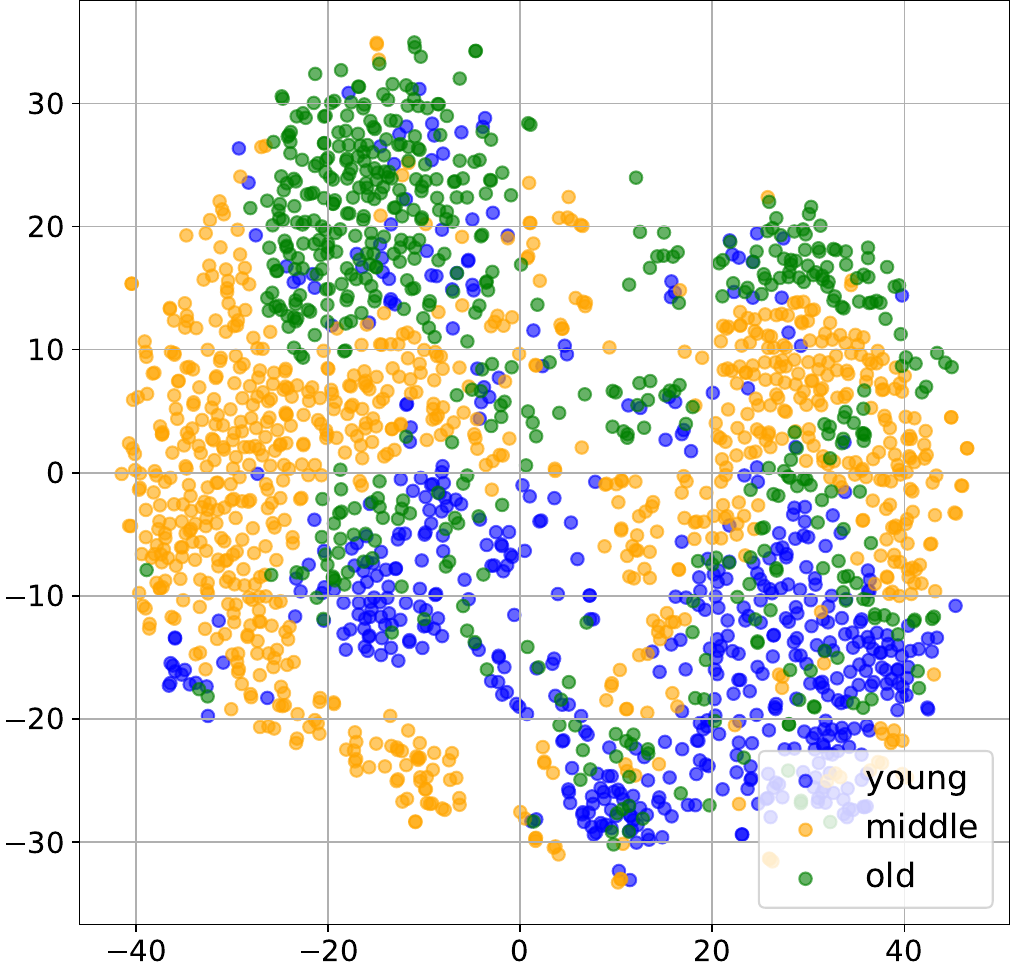} 
        \caption{Wav2Vec  generated dataset}
        \label{fig:3b}
    \end{subfigure}

    \caption{ t-SNE distributions of different features from original and generated datasets.}
    \label{fig:t-sne}
\end{figure}

\section{Experiments}
\label{sec:experiment}
% discuss result 
For the age conversion into a neutral, middle-aged age domain, we  train two generators (\(G_{y2m}\) and \(G_{o2m}\)) and corresponding discriminators (\(D_{y2m}\) and \(D_{o2m}\)) to convert young and old audio into middle-aged audio while leaving the data originally in the middle-aged domain untouched. The generated Mel-spectrograms are subsequently converted back into speech using a pre-trained HiFi-GAN vocoder \cite{kong2020hifi}. The length of each generated speech segment is fixed to 3 seconds. Subsequently, the three feature types (i-vector, x-vector, and Wav2Vec) are extracted from both the original dataset and the dataset generated by the age conversion model. The features are then mapped to a low-dimensional space through t-distributed stochastic neighbour embedding (t-SNE) to observe their distribution patterns, similar to~\cite{baird2021prototypical}.

Figure \ref{fig:t-sne} illustrates the t-SNE representation results. The overall distribution of the generated data appears more compact compared to the original dataset. This suggests that the generated data has reduced the influence of age-domain bias to some extent, resulting in samples with greater similarity and more consistent converted speech.

Table \ref{table:acc} presents the AKV overall weighted accuracy across different combinations of models and features. Additionally, it reports the accuracy of various kinship relationships under different feature and model combinations. The results represent the average of five independent experiments. We first discover that the optimised TripletNet model considerably improved accuracy, surpassing the Baseline TripletNet across the original and generated datasets. Second, the introduced Wav2Vec feature demonstrated superior performance, outperforming traditional i-vectors and x-vectors in almost all kinship relation combinations. Additionally, experiments in the generated dataset achieved higher accuracy across several kinship relations. In generated dataset, The overall weighted accuracy in Wav2Vec features with optimised TripletNet reached the highest score, with 71.3\%, representing an improvement of approximately 5\% over the baseline. Notably, for typical cross-age relationships such as father and daughter or mother and daughter, Wav2Vec achieved accuracy of 74.0\% and 71.1\%, respectively, on the generated dataset, highlighting its strength in handling age differences. This demonstrates that the generated data enhanced the model's generalisation ability and effectively mitigated the negative impact of age-domain bias.

\section{Conclusion}
\label{sec:conclusion}
In this paper, we proposed a voice conversion-based kinship verification method. This method mitigated the domain bias caused by age deviation by projecting all audio into the age-standardised domain. We employed the CycleGAN-VC3 architecture to perform age domain conversion of audio and generated a new data set. In addition, we optimised the metric learning method in the baseline and introduced a new feature extraction method based on Wav2Vec 2.0 into the task. The experimental results demonstrated that the method clearly improved the accuracy of audio-based kinship verification task. Future work could consider additional gender conversion to better cope with tasks such as brother and sister, father and daughter, or mother and son.

\bibliographystyle{IEEEtran}
\bibliography{refs}

\end{document}